# Coupling Dichroism in Strong-Coupled Chiral Molecule-Plasmon Nanoparticle System

Nan Gao[1, #], Haoran Liu[1, #], Yurui Fang[1, *]

[1.] School of Physics, Dalian University of Technology, Dalian 116024, P.R. China.

* Corresponding authors: yrfang@dlut.edu.cn (Y.F.)

# These authors contributed equally.

**Abstract**

The interaction between intense light-matter not only promotes emerging applications in quantum and nonlinear optics but also facilitates changes in material properties. Plasmons can significantly enhance not only molecular chirality but also the coupling strength. In this study, we investigate the coupling dichroism in a strongly coupled chiral molecule–plasmonic nanoparticle system using RT-TDDFT. By simulating the interaction between L/D- Phenylglycinol molecules and chiral aluminum clusters (Na-doped $Al_{197}Na_4$), we examine the effects of molecular chirality, cluster chirality, and the coupled effect in the system. Our results demonstrate that the achiral/chiral clusters induce significant spectral shifts and enhance molecular CD signals due to strong plasmon-molecule coupling. The electric-field distribution and transition contribution maps (TCMs) reveal the formation of bonding and antibonding polaritonic modes, modulated by molecular proximity to the cluster. Both of the coupling factor and decay rate of the coupled system will be modulated by the chirality of the molecules and the cluster. Furthermore, we find that increasing the number of coupled molecules leads to a substantial increase in the intensity of lower polaritonic modes, highlighting the collective behavior in multi-molecule systems due to the modal crosstalk or resonance between cluster chirality and molecular chirality. These findings provide valuable insights into the fundamental mechanisms governing plasmon-enhanced chirality at the atomic scale, which have implications for the design of highly sensitive chiral sensors and optoelectronic devices.

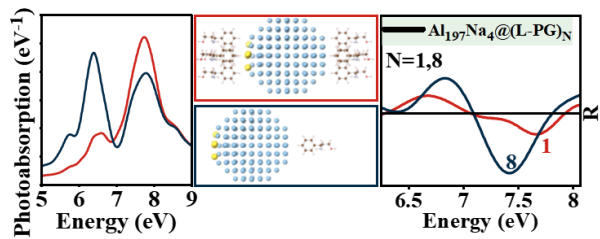

TOC

**Introduction**

The optical properties of all materials stem from their quantum mechanical response to perturbed light. As nanoparticle (NP) sizes shrink below 10 nanometers (nm), the absorption spectra exhibit the characteristic features of quantum dots with discrete transitions between quantum energy levels[1,2]. In the size range of 1 to 5 nm, traditional classical electromagnetic theories fail to fully account for quantum effects, as the plasmonic resonance peaks do not shift with size, thus not predicting size dependence[3]. Currently, a variety of semi-classical models have been employed to delineate the plasmonic resonances at the quantum scale[4–7]. Nevertheless, it is essential to incorporate the influences of quantum mechanics when comprehensively modeling plasmonic NPs. Moreover, the Jellium model does not consider the significance of d-electrons and atomic structure in noble metals (such as Cu, Ag, Au, etc.)[8,9], which are indispensable for accurately describing noble metal NPs. Exact ab initio computational methods are mainly applicable to small clusters: time-dependent density functional theory (TD-DFT)[10-12] has been successfully applied to simulate small clusters like $Ag_{55}$, $Ag_{147}$[13], and even nanoshell structures up to the size of $Ag_{272}$[14]. This demonstrates that a fully quantum mechanical treatment is desirable for predicting the properties of nanoscale systems, and TD-DFT is an effective tool for analyzing electronic excitation properties. Recently, researchers have used real-time propagation TD-DFT (RT-TDDFT) methods to simulate the strong coupling phenomena between metal clusters of 3-5 nanometers in size and small molecules[15–18]. Furthermore, the study of strongly coupled systems has also extended into the field of hot electrons[19–22]. However, the aforementioned research is almost exclusively limited to non-chiral systems, with few reports on plasmonic enhancement of chirality at the atomic level[23].

Chirality plays a pivotal role in biological systems: for example, the spiral structure of DNA is right-handed, whereas amino acids are commonly left-handed. In the realm of pharmaceutical development, chirality is equally significant. Thalidomide, a drug widely used to treat morning sickness during pregnancy, possesses chiral properties, and its chiral isomers have severe teratogenic effects[24]. The two enantiomers of a molecule can dictate whether a drug will yield the anticipated beneficial outcomes or become toxic. Hence, it is crucial to identify the optical activity of chiral systems. Chiral molecules exhibit distinct absorption characteristics for circularly polarized light with varying rotations, a phenomenon that can be detected through circular dichroism (CD) spectroscopy. The CD is calculated as $\Delta\varepsilon = \varepsilon_L - \varepsilon_R$, where $\varepsilon_L$ and $\varepsilon_R$ represent the extinction coefficients for left- and right-circularly polarized light, respectively. CD technology is a powerful tool for characterizing chiral systems and distinguishing enantiomers due to its high sensitivity to minute changes in molecular structure and its uniqueness to each conformation[25]. Structural, dynamic, and thermodynamic information can be obtained through CD measurements. The opposite nature of chiral molecular CD spectral signals can be utilized to discern the optical rotation properties of molecules. The CD spectra of organic molecules are usually weak, primarily manifesting in the ultraviolet region, which limits their application in the field of molecular chirality sensors[26].

Plasmonic NPs can effectively enhance molecular CD signals[27,28]. Various experiments have shown that plasmonic NPs can amplify the magnitude of molecular CD signals and red-shift their optical characteristics to the plasmonic band[29,30]. The plasmonic enhancement of CD can currently be explained by theoretical models based on plasmon-molecule dipole and multipole Coulomb interactions[31,32], or by numerical solutions of electromagnetic equations[33,34]. Currently, chiral light-matter interactions have been explained as a combination of five distinct contributions: non-resonant interactions, changes in mode excitation and emission efficiency, modal resonance shifts, and intermodal crosstalk[33]. However, in past studies, the descriptions of molecules and NPs have been simplified and rely on preset parameters. Therefore, these theoretical models are unable to explain the dependencies of plasmonic enhancement at the atomic level, necessitating rigorous ab initio calculations to bridge theory and experiment. There has been work analyzing nanoscale metal-organic clusters using the LCAO-RT-TDDFT method[35]. Ville J. Härkönen and colleagues have systematically investigated the CD enhancement of molecules on silver NPs or sandwiched between two silver NPs using TDDFT, as well as the influence of molecular orientation, molecular-NP gap, and molecular coverage on CD enhancement[23]. This provides valuable guidance for a deeper understanding of the intrinsic mechanisms of plasmonic enhancement of chirality. However, the intrinsic mechanism of plasmonic enhancement of chirality in strongly coupled light-matter interaction systems remains unclear. Here, we select $Al_{201}$ clusters coupled with L/D-Phenylglycinol (PG) chiral molecules to examine the enhancement mechanism of plasmons on left-handed and right-handed chiral molecules, considering the gap and the number of coupled molecules as well.

In this paper, we employ RT-TDDFT to investigate plasmonically enhanced molecular CD and the coupling dichroism in strongly coupled systems at the atomic level. We precisely simulate the CD enhancement effects, taking into account the intrinsic chirality of the NP, the orientation of molecular chirality, the gap between the NP-molecule, as well as the coupling of multiple molecules to a single chiral NP. We provide an exact simulation of the electronic interactions within the chiral molecule-NP hybrid system. The matching and mismatching of the structure chirality for the molecule and the cluster will significantly influents the coupling strength, Rabi splitting energy and decay rate. As the number of molecules increases, collective effect will appear for the molecules through the intermedium of cluster. In previous studies, plasmonic clusters have often been assumed to be non-chiral, which is disadvantageous for a full understanding of chiral mechanisms. Our research includes the exploration of the impact of chiral clusters on the overall chiral properties of the system, our results figure out the coupling dichroism and collective effect in strong coupled plasmon-exciton system. We hope this provides a valuable reference for further understanding chiral mechanisms at the atomic level.

**Computational Details.** In this study, we performed (TD)DFT[36] calculations using the GPAW package[37] to achieve structural relaxation of metal clusters and individual molecules. The calculations employed the Perdew-Burke-Ernzerhof (PBE)[38] generalized gradient approximation as the exchange-correlation functional. The calculations were carried out in the linear combination of atomic orbitals (LCAO) mode[39], with a grid spacing of 0.3 Å, and the number of empty bands adjusted based on the number of occupied orbitals to meet subsequent computational needs. To address finite temperature effects, a Fermi-Dirac smearing of 0.05 eV was implemented. For the structural optimization of all individual metal NPs and molecules, the Broyden-Fletcher-Goldfarb-Shanno (BFGS) quasi-Newton algorithm[40,41] was applied through the Atomic Simulation Environment (ASE)[42] interface. The optimization was carried out until the maximum force on any atom was below 0.05 eV/Å. Subsequent calculations utilized these already optimized structures without re-optimization, which is reasonable and sufficient for most computations. Based on this, we simulated the ground-state electronic structure properties of all monomers and complex configurations, selecting the default projector augmented-wave data sets[43] and appropriate dzp or pvalence basis sets[44] for different elements. Using the Direct-LCAO method as the eigensolver, setting the density convergence criterion to $1\times 10^{-12}$, and combining the Mixer mixing technique with the Fermi-Dirac distribution model, we ensured the convergence and stability of the calculations. Additionally, a time limiter was introduced to effectively manage computational resources, and the maximum number of iterations was set to 500. In addition, the LCAO-RT-TDDFT implementation[45] in GPAW was used for the RT-TDDFT calculations. A δ-kick strength of $K_z=10^{-5}$ in atomic units was used. The time propagation was done in steps of 10 attoseconds (as) for a total length of 30 femtoseconds (fs) using the adiabatic PBE kernel. The ASE library was used for constructing and manipulating atomic structures. The NumPy[46], SciPy[47], and Matplotlib[48] Python packages, the VMD software[49] and VESTA[50] were used for processing and plotting data. The emcee[51] Python package used to fit spectra to the coupled oscillator model.

## Results and discussion

**Geometric Structure of Coupled Chiral Clusters and Chiral molecules.** We utilized $Al_{201}$ cluster as the primary plasmonic structure, which can generate strong and broad plasmonic dipole mode absorption peaks (approximately 7.7 eV), as evidenced in previous literature[15]. To endow chirality, we doped four sodium atoms into the {100} facets of the $Al_{201}$ cluster, arranging these atoms into the spatial pattern of an L-shaped (see Figure 1, illustration on the right). We have designated the doped chiral cluster as $Al_{197}Na_4$. This structure exhibits pronounced chiral spectral features (see Figure 2c), and its optical absorption spectrum is relatively weaker yet broader than the undoped structure, while the peak position is conserved at approximately 7.7 eV, similar to the $Al_{201}$ cluster. To achieve strong interaction and chiral response, we selected the chiral molecule L/D-Phenylglycinol (PG)[52,53], whose peak positions are close to that of the plasmonic cluster. The structures of these molecules are shown in Figure 1a. To investigate the influence of gap between the plasmonic cluster and the chiral molecules for coupling and chiral properties, we aligned the phenyl ring functional group of the chiral molecules directly towards the {100} facets of the cluster at varying gap $d$, where $d$ was set to 2-8, 15, 25, and 50 angstroms (Å). For comparative purposes, we designed four configurations: $Al_{201}$@L-PG, $Al_{201}$@D-PG, $Al_{197}Na_4$@L-PG, and $Al_{197}Na_4$@D-PG. Their structures are depicted in Figure 1b.

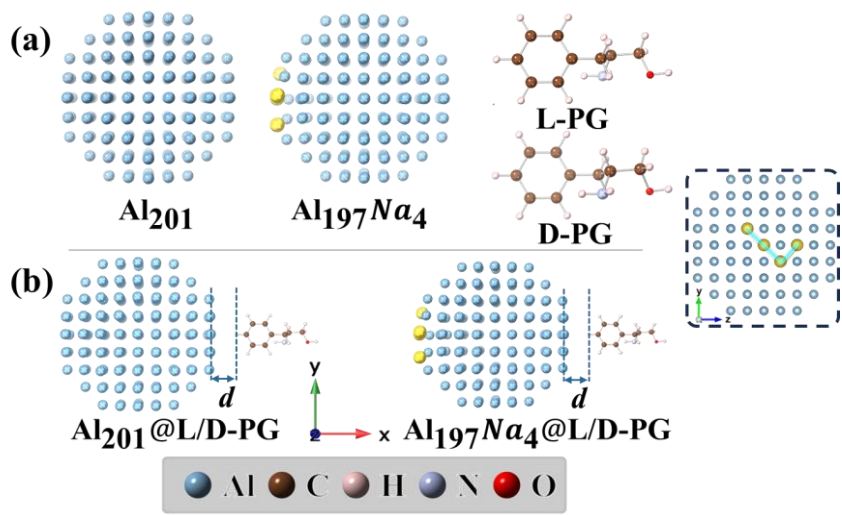

Figure 1. The geometric structures of monomers and complexes. (a) the geometric structure of $Al_{201}$, $Al_{197}Na_4$, L-PG, and D-PG, respectively. (b) the cluster complex chiral molecular system of $Al_{201}$@L/D-PG and $Al_{197}Na_4$@L/D-PG, where gap between clusters and single molecules is set to 2-8, 15, 25, and 50 Å. The illustration on the right shows Na atoms doped in an L-shaped spatial configuration.

**Excitation in Strong Coupled Monomers and Complexes with Chirality.** Now, considered a plasmonic system based on an Al NP as proof of concept. Compared to noble metals with d-electron screening plasmonic, the nearly free-electron electronic structure of Al greatly simplifies the analysis. Moreover, Al has garnered recent interest as an alternative plasmonic material[15,17,18]. For plasmonic nanoclusters of Al metals, light absorption primarily occurs in the ultraviolet region. We first examined the spectral changes with cluster-molecule gap for systems composed of $Al_{201}$@L/D-PG. We performed RT-TDDFT simulations on the entire system to capture not only a wide range of couplings, but also phenomena such as charge transfer and the modulation of excited states arising from NP-molecular interactions. As shown in Figure 2a, for the $Al_{201}$@L/D-PG, the photoabsorption spectrum exhibits a strong and broad absorption peak at ~7.7 eV, with the absorption intensity of the chiral cluster being lower than that of the non-chiral cluster (Figure 2c). As the $Al_{201}$ NP and the L/D-PG molecule approach each other, they resonantly couple to form two polaritons. In the transition contribution map (TCM) of the coupled system (refer to Figure 3 and Supplementary Notes S1 in the *Supporting Information*), the lower (6.55 eV) and upper (7.69 eV) polaritons correspond to the symmetric and antisymmetric plasmon-molecule hybrid states, respectively. The TCM visually decomposes the energy of individual electron-hole (e-h) transition contributions and informs about the interactions between e-h pairs. By comparing with the TCM of the uncoupled components, the individual contributions to the polaritons are revealed. Low-energy transitions (≲2 eV) near the Fermi level form the cluster plasmons, while the contributions around -2 to 3 eV originate from the molecular excitons. Crucially, the two polariton states in Figure 3 differ in the sign of the molecular contribution. For the lower polariton (LP), the plasmonic and molecular transitions are in phase, whereas for the upper polariton (UP), they are out of phase. This corresponds to symmetric and antisymmetric combinations, respectively, and is also clearly visible in the induced density. At the LP, the dipoles of the particle and the molecule are parallel, while at the UP, the dipoles of the particle and the molecule are antiparallel. This arrangement is prototypical of a strongly coupled system. This observation holds true for the other systems discussed below, indicating that strong or near-strong coupling is already present at the single molecule gap of 2 Å from the $Al_{201}$ NP. Here we note that for the lower polariton, the induced density of the molecular unit in the coupled system does not show a significant difference from the uncoupled one. However, for the upper polariton mode, the induced density on the side of the molecule closest to the cluster undergoes a phase reversal, which precisely forms an antibonding mode of a coupled system, whereas for the lower polariton, it is precisely a low-energy bonding mode.

As the gap increases, we observe that the intensity of the LP peak in the photoabsorption spectrum gradually diminishes, which is closely related to the decrease in coupling strength. In this process, the molecular contribution to the UP mode progressively fades, while for the LP mode, the primary contribution is dominated by the molecule, with the cluster's contribution correspondingly diminishing. This can be confirmed by the TCM and induced density plots at Figure S1 in the *Supporting Information*. When the gap is sufficiently large (15 Å, 25 Å, and 50 Å in Figure 2a), we find that the photoabsorption spectra of these coupled systems nearly coincide with the spectrum of the $Al_{201}$ cluster, there is no interaction between the molecule and the cluster, which verifies that the small gap system upper polariton is indeed a result of the strong coupling effect of the interaction, rather than a simple superposition of the spectra of the molecule and the cluster.

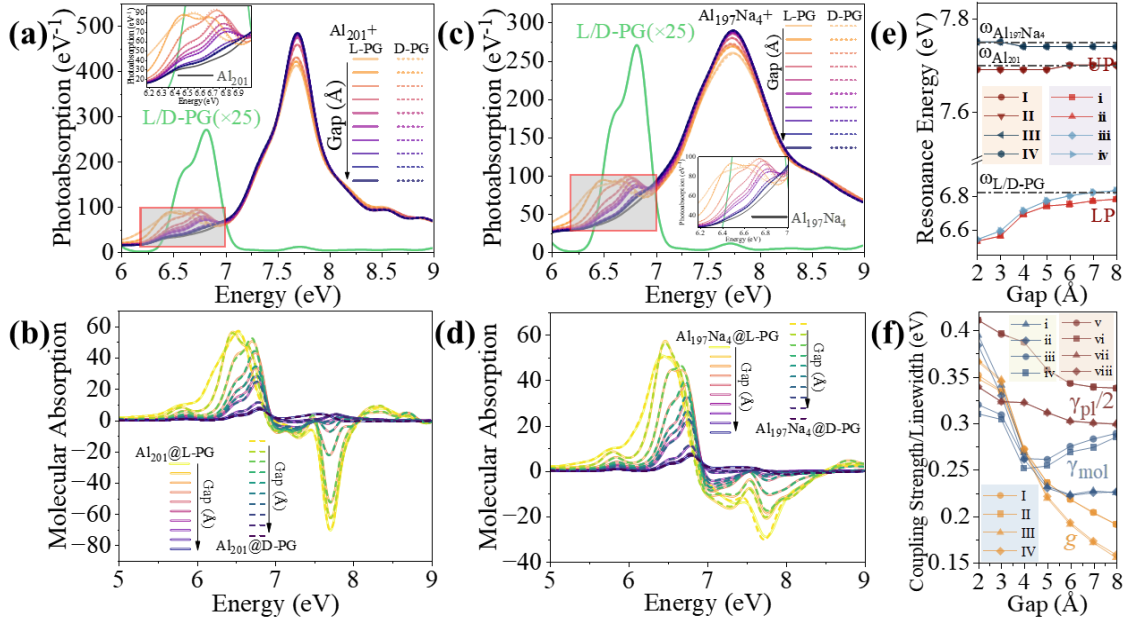

Figure 2. Photoabsorption spectrum and coupling strength. (a) is the photoabsorption for $Al_{201}$@L/D-PG with different gap (kick is x), where illustration shows a magnified section of spectrum. (b) molecular absorption spectra with different gaps reveal in-phase and out-of-phase polaritons, as well as their increased separation and broadening for decreasing separation for $Al_{201}$@L/D-PG. (c) the photoabsorption for $Al_{197}Na_4$@L/D-PG with different gap (kick is x), where illustration shows a magnified section of spectrum. (d) the molecular absorption for $Al_{197}Na_4$@L/D-PG. (e) is the resonance energy of $Al_{201}$@L/D-PG and $Al_{197}Na_4$@L/D-PG with different gap, where I-IV represents the UP of $Al_{201}$@L-PG, $Al_{201}$@D-PG, $Al_{197}Na_4$@L-PG, and $Al_{197}Na_4$@D-PG, respectively, i-iv represents the LP of $Al_{201}$@L-PG, $Al_{201}$@D-PG, $Al_{197}Na_4$@L-PG, and $Al_{197}Na_4$@D-PG, respectively. (f) the coupling strength and linewidth of $Al_{201}$@L/D-PG and $Al_{197}Na_4$@L/D-PG with different gap, where I-IV represents the $g$, i-iv represents the linewidth of LP, and v-viii represents the linewidth of UP for $Al_{201}$@L-PG, $Al_{201}$@D-PG, $Al_{197}Na_4$@L-PG, and $Al_{197}Na_4$@D-PG, respectively.

In addition, we also consider the system of chiral cluster coupled with chiral molecules ($Al_{197}Na_4$@L/D-PG). Owing to the incorporation of Na atoms, the $Al_{197}Na_4$ cluster manifests an asymmetric dipole mode. The free electrons in the Na atoms are more readily excited, resulting in a greater non-locality of the s electrons compared to the p electrons. This leads to a non-homogeneous distribution of the induced density, which constitutes the origin of its chirality (see Figure 3 and Figures 4c-d). In the coupled state, the photoabsorption spectral characteristics of $Al_{197}Na_4$@L/D-PG resemble those observed in $Al_{201}$@L/D-PG, with the sole distinction being a slightly weaker intensity of the UP peak compared to $Al_{201}$@L/D-PG. When the gap between the two is sufficiently close (2 Å), the lower polariton forms a low-energy bonding mode, while the upper polariton constitutes a high-energy antibonding mode (see Figure 3). As the gap increases gradually, the interaction weakens correspondingly, with the cluster's contribution to light absorption in the LP mode diminishing and increasingly dominated by the molecule. In the UP mode, the molecule's contribution to light absorption diminishes and eventually disappears, leaving the cluster's contribution as the predominant factor (refer to Figure S2 in the *Supporting Information*). This is consistent with the observations in $Al_{201}$@L/D-PG. Similarly, when the gap is sufficiently large (15 Å, 25 Å, and 50 Å in Figure 2c), the spectral lines of the photoabsorption spectra have completely overlapped with those of $Al_{197}Na_4$, exhibiting behavior identical to that observed in $Al_{201}$@L-PG. This indicates that at these gaps, the interaction has completely vanished, thereby validating that the photoabsorption spectra at gaps less than 8 Å are the result of strong coupling, rather than a simple linear superposition of spectra. For theoretical calculations, the absorption spectra of left-handed and right-handed molecules completely overlap, which is reasonable. For the composite structures, a comparison reveals that the photoabsorption spectra of X@L/D-PG (where X represent $Al_{201}$ or $Al_{197}Na_4$) remain consistent (compare the solid and dashed lines in Figures 2a or 2c.), and their charge transfer information corroborates this finding (refer to Figure 3 and Figures S1-2 in the *Supporting Information*).

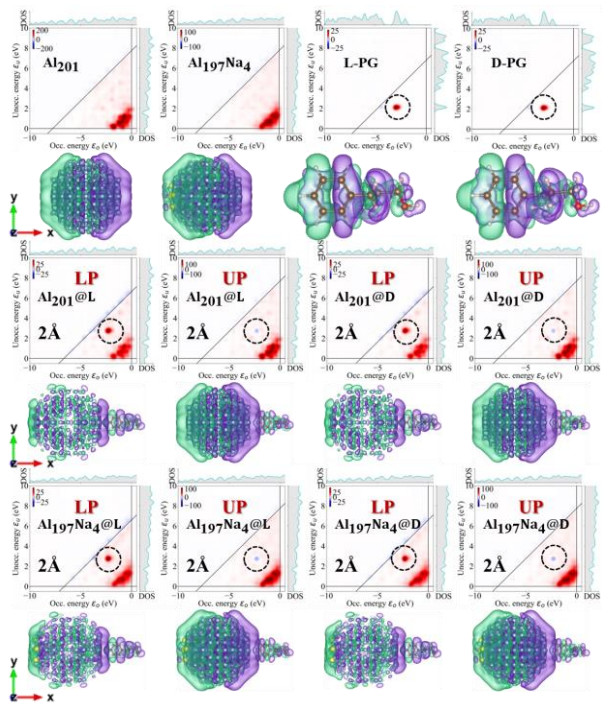

Figure 3. The TCM and induced density for $Al_{201}$, $Al_{197}Na_4$, L-PG, D-PG, $Al_{201}$@L-PG, $Al_{201}$@D-PG, $Al_{197}Na_4$@L-PG, and $Al_{197}Na_4$@D-PG along with the density of states (DOS), respectively. For complex structures, the gap between molecular and cluster is 2 Å. Where L and D represent L-PG and D-PG, respectively.

Furthermore, in the complete spectrum of $Al_{201}$@L/D-PG depicted in Figure 2b, the pure molecular absorption changes caused by different gaps can be observed in detail. As the molecule approaches the NP, a broadening of the molecular absorption band is evident, which aligns with the predictions of molecular resonator absorption band broadening based on the macroscopic coupled oscillator theory[15]. At the energy levels corresponding to the UP branch, the absorption exhibits a negative effect, which is consistent with the emergence of an anti-symmetric dipole configuration. Conversely, at lower polariton energies, a symmetric dipole arrangement is observed, with the molecular absorption being positive. A faint negative light absorption can already be observed at an 8 Å gap, indicating that relatively weak coupling still exists at this distance. However, as the gap increases beyond 15 Å, the negative contribution vanishes, suggesting that coupling is no longer present at sufficiently large gaps. Therefore, the conventional criterion for the strong coupling state (i.e., $g > \sqrt{(\gamma_{ex}^2 + \gamma_{pl}^2)/2}$ )[54] can only be applied to the fitting of coupled oscillators when a gap of ≤8 Å is satisfied. Figure 2d shows the pure molecular absorption changes in $Al_{197}Na_4$@L/D-PG caused by different gaps. Compared to $Al_{201}$@L/D-PG, the negative absorption is lower and multiple peaks are present, indicating that the presence of chiral cluster has a significant impact on strong coupling at distances below 8 Å. This influence will be further explored in the context of chirality enhancement in the following sections.

**Gap Dependent Coupling Strength with Chirality.** The velocity-coupled harmonic oscillator framework serves as a potent tool for elucidating both the photoabsorption characteristics and the extent of coupling within NP-cluster systems. To quantitatively analyze the spectral features of these hybrid systems, we employ a sophisticated coupled oscillator model. This model allows for the precise fitting of the absorption spectra, thereby providing insights into the interplay between NP and molecular resonances. The spectra $\sigma_{\text{abs}}(\omega)$ of coupled NP with molecular are fitted by[18,55]

$$\sigma_{\text{abs}}(\omega) = a\omega \operatorname{Im}[L_{ex}(\omega)/(L_{ex}(\omega)L_{pl}(\omega) - 4g^2\omega^2)] \qquad (1.1)$$

where $a$ is an amplitude, $L_x(\omega) = \omega_x^2 - (\omega + i\gamma_x/2)^2$ are the complex form of the Lorentz function corresponding to the exciton (often is molecular, $x = ex$) and the plasmonic ($x = pl$) oscillators, and $\omega_{ex}$ and $\gamma_{ex}$ are the exciton resonance frequency and width, $\omega_{pl}$ and $\gamma_{pl}$ is the plasmon resonance frequency and width. The coupling strength between the oscillators is $g$. The parameters of resonance frequency, width and coupling strength $g$ for $Al_{201}$@L/D-PG and $Al_{197}Na_4$@L/D-PG are plotted in Figures 2e-f (the relevant data are listed in Tables S1-S4 in the *Supporting Information*). The spectra obtained from fitting with the coupled oscillator model are consistent with the spectra obtained from TDDFT calculations across the entire range of gaps considered (refer to Figure S4 in the *Supporting Information*).

As shown in Figure 2e, In the $Al_{201}$@L/D-PG system, the resonance energy of the UP mode shows a minimal deviation from the plasmonic resonance energy of the isolated $Al_{201}$ cluster, exhibiting a slight blue-shift with increasing gap. In contrast, for the $Al_{197}Na_4$@L/D-PG system, the resonance energy of the UP mode slightly red-shifts as the gap increases. Overall, the plasmonic resonance energies of both systems are similar to those of the isolated plasmonic clusters, indicating that the resonance peaks are not influenced by the coupled molecules. However, for the LP mode of the coupled systems, a pronounced blue-shift in resonance energy is observed with increasing gap, with the $Al_{197}Na_4$@L/D-PG system exhibiting a stronger blue-shift than the $Al_{201}$@L/D-PG system. When the gap increases to 8 Å, the resonance energy of the LP mode gradually aligns with the peak value of the single molecule, verifying that at this point, the interaction is weakened or even negligible due to the sufficient distance between the cluster and the molecule. Consequently, the LP mode is predominantly governed by the absorption of the single molecule, which is consistent with the TCM diagrams showing the molecular contribution to the UP mode is almost negligible and the plasmonic contribution to the LP mode is sufficiently weak at a large enough distance.

Figure 2f summarizes the resonance peak broadening and coupling strength ($g$) for the $Al_{201}$@L/D-PG and $Al_{197}Na_4$@L/D-PG systems as the gap increases. Clearly, as the gap increases, leading to a weakening in interaction, the coupling strength correspondingly decreases. Interestingly, coupling the same cluster to a chiral molecule or its enantiomer does not significantly affect the coupling strength, which aligns with the trends observed in the photoabsorption spectra. However, the coupling strength of chiral cluster-molecule systems is slightly lower than that of non-chiral cluster-molecule systems at larger gaps, which can be attributed to the reduced difference in absorption strengths between the molecule and the cluster in the $Al_{197}Na_4$@L/D-PG system, a factor that is favorable for the $g$. Additionally, the fitting width indicates a significant increase in the oscillator width, reaching a maximum at a 2 Å gap. Notably, at the largest gap, as $g$ trends toward 0, the molecular resonance is no longer driven and undetectable within the framework of the current coupled oscillator model. Furthermore, with increasing gaps, the plasmonic oscillator converges towards that of the isolated cluster.

In addition, the coupling strength should also be influenced by the atomic doping of the cluster. For instance, the $Al_{201}$ cluster has a stronger electric field compared to the doped chiral cluster $Al_{197}Na_4$, where the doped *Na* atoms can more effectively localize the electric field in the vicinity (Figure S5 in *Supporting Information*). It is well known that when the refractive index of a material is higher than that of the surrounding environment, it leads to additional concentration of electromagnetic fields, thereby enhancing the coupling efficiency. Although the refractive index cannot be used to accurately describe a single chiral molecule, it can similarly produce an enhanced effect. This is supported by the data on electric near-field enhancement based on RT-TDDFT calculations (see Figures S5 a-b in *Supporting Information*). In the LP mode of the coupled system, compared to the plasmonic field of the bare cluster, the chiral molecule can better focus the electric field, thereby adjusting the characteristics of the cavity and altering the field distribution in vacuum. For the UP mode of the coupled system, the chiral molecule located within the hot spot of the cluster's plasmonic field promotes the molecule's field, which also confirms the negative absorption of the molecule in the coupled system (Figures 2b and d). However, as the gap increases, the electric field of the molecule becomes significantly weaker. Due to the inherently weaker electric field of the doped chiral cluster, the coupling strength in systems with the $Al_{197}Na_4$ cluster gradually becomes weaker than that of the hybrid system with the $Al_{201}$ cluster as the gap increases (Figure 2f). Next, we delve deeper into the relationship between the system's chirality and coupling strength through rotor strength ($R$) analysis.

**Gap Dependent Enhanced Chirality.** Plasmonic can effectively enhance molecular chirality, making it straightforward to predict the electronic circular dichroism (ECD) of systems using first-principles calculations. Within the LCAO-RT-TDDFT framework implemented in the GPAW package, ECD can be computed from the time-dependent magnetic dipole moments[35]. Although the molecular features might be obscured by the intense plasmonic absorption of NPs[23], they remain clearly discernible in the CD spectra, as illustrated in Figures 4a-d. Consequently, measuring the CD of molecules adsorbed on metal NPs is a promising approach for detecting molecular chirality. Although the CD enhancement in coupled systems can encompass multiple processes, and the definition of enhancement factors might be nebulous, in our system, the molecular CD signatures appear as anticipated within the energy range of interest (~6-8 eV) and exhibit trends that are straightforward to analyze. Therefore, we focus on the enhancement of the two lowest-energy peaks (I and II) within the rotary strength energy range. Considering the presence of the chiral cluster, for the purpose of analysis, we uniformly calculated the rotary strengths of isolated chiral/non-chiral clusters and then performed a subtraction process. Thus, we define the enhancement chirality factor $\mathscr{P}^{\mathbf{I/II}}$ as:

$$\mathscr{P}^{\mathbf{I/II}} \equiv [R_{tot} - R_{clu}]^{\mathbf{I/II}}(\omega') \Big/ R_{mol}^{\mathbf{I/II}}(\omega'') \tag{1.2}$$

where $R_{tot} - R_{clu}$ represents the difference between the total rotary strengths and those of the isolated cluster, indicating the enhanced molecular rotary strength spectrum. $[R_{tot} - R_{clu}]^{\mathbf{I/II}}(\omega')$ and $R_{mol}^{\mathbf{I/II}}(\omega'')$ correspond to the rotary strengths of the enhanced molecule and the isolated molecule at the first positive (negative) peak or the second negative (positive) peak, respectively. It is important to note that the peak energies $\omega'$ and $\omega''$ do not necessarily remain the same. All detailed parameters are listed in Tables S7-10. As shown in Figures 4a-b, the isolated molecules exhibit completely opposite chiral signals (the total signals are shown in Figure S3). Figure 4a shows the rotary strength of L-PG on the side of $Al_{201}$ cluster for different gaps. The maximum enhancement occurs when the gap is 2 Å (with $\mathscr{P}^{\mathbf{I}} = 3.38$ and $\mathscr{P}^{\mathbf{II}} = 5.41$), and the enhancement factor gradually decreases as the gap increases, reaching a minimum value when the gap is 8 Å (with $\mathscr{P}^{\mathbf{I}} = 1.77$ and $\mathscr{P}^{\mathbf{II}} = 3.01$) (Figures 4e-f). The position of the molecule relative to the cluster alters the rotary strength of the coupled system by approximately a factor of 2 and can either enhance or diminish the chiral photonic signal of the molecule. Figure 4b displays the rotary strength of D-PG on the side of $Al_{201}$ cluster for different gaps. In comparison to the $Al_{201}$@L-PG, the two peaks of interest are exactly opposite. The most significant enhancement is observed when the molecule is in closest proximity to the cluster (with $\mathscr{P}^{\mathbf{I}} = 3.66$ and $\mathscr{P}^{\mathbf{II}} = 5.64$), with the enhancement diminishing progressively as the gap increases. This is accompanied by a roughly twofold alteration in the rotary strength, as depicted in Figures 4e-f. The plasmonic can effectively enhance the CD signals of molecules located in regions of strong interaction hotspots. For the two molecules that are enantiomers, both are enhanced, and the magnitude of the enhancement is the same. The enhanced CD signals maintain the same good opposite nature as the original CD signals. In the case of $Al_{197}Na_4$@L-PG (Figure 4c), the inherent chirality of the cluster, which is an order of magnitude stronger than that of the molecule, leads to a relatively complex pattern of chirality enhancement. When the gap is 2 Å, the maximum enhancements are $\mathscr{P}^{\mathbf{I}} = 5.90$ and $\mathscr{P}^{\mathbf{II}} = 7.56$, respectively, marked increases compared to those of $Al_{201}$@L-PG. As the gap increases, the enhancement gradually decreases. At a separation of 8 Å, the enhancements bottom out at $\mathscr{P}^{\mathbf{I}} = 1.90$ and $\mathscr{P}^{\mathbf{II}} = 3.54$, respectively. At this point, the enhancement factors align more closely with those of $Al_{201}$@L-PG (Figures 4e-f). At such a gap, the interaction has almost vanished, and whether the cluster is chiral or not has a minimal effect on the mechanism by which molecular chirality is enhanced. At the smallest gap, the interaction is robust, and at the juncture not only does the localized surface plasmonic resonance (LSPR) of the cluster contribute to the enhancement of molecular chirality, but the cluster's own chiral characteristics also positively influence the augmentation of molecular chirality. Figure 4d shows the rotary spectra of $Al_{197}Na_4$@D-PG with different gaps, which are roughly opposite to those of $Al_{197}Na_4$@L-PG, but the line shapes are significantly different. For the enhancement of the first peak, the relationship between distance and enhancement is not linear; as the distance increases, the enhancement first increases and then decreases. For the second peak, the enhancement is inversely related to the increase in distance, which is consistent with $Al_{197}Na_4$@L-PG. The coupling of the chiral cluster with L-PG or D-PG affects the enhancement of chirality, indicating that when both components of the coupled system are chiral, the chirality enantiomers of the molecule will also produce different chiral responses.

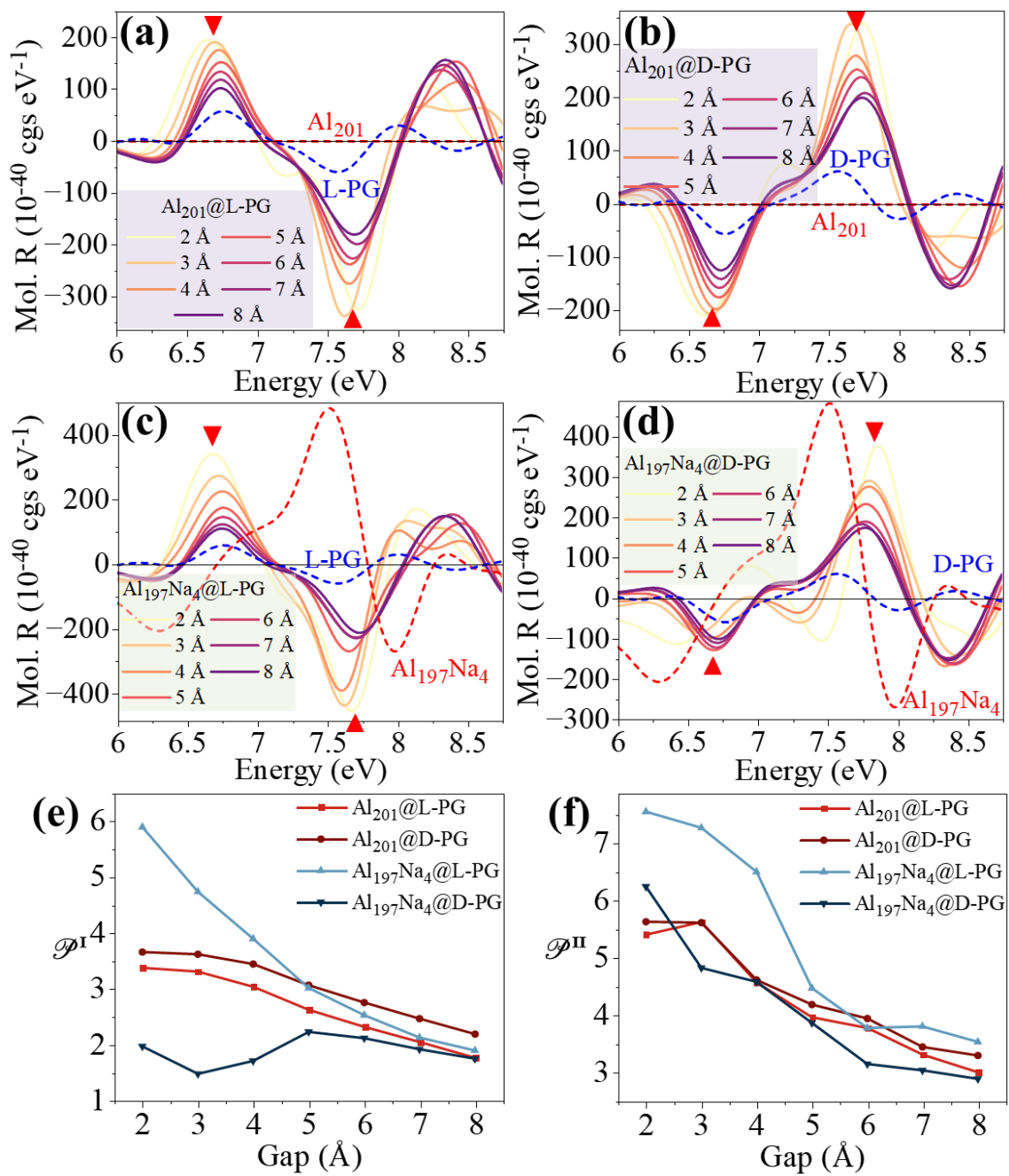

Figure 4. The plasmon enhanced CD spectra. (a-d) the enhancement molecular CD for $Al_{201}$@L-PG, $Al_{201}$@D-PG, $Al_{197}Na_4$@L-PG, and $Al_{197}Na_4$@D-PG with different gap, respectively. The red triangle in the figure marks the position of the peak of interest. (e-f) is the parameter of enhancement chiral factor.

**Enhanced Strong Coupling by Collective Excitation of Molecules.** To consider the contribution of the number of molecules on the coupling strength and the impact of chirality, the chiral molecules are placed at two opposing {100} facets of the clusters (Figure 5) and no additional structural relaxation was performed on the coupled system. The structure in Figure 5a is $Al_{201}@(L\text{-}PG)_N$, while the structure in Figure 5b represents the coupling configuration of the chiral cluster with left-handed molecules [$Al_{197}Na_4@(L\text{-}PG)_N$]. Similarly, we have also designed and investigated the coupling structures of different numbers of D-PG with the plasmonic clusters. Their configurations are similar, with the only difference being the replacement of L-PG with D-PG. For the coupling structures with varying numbers of molecules, we maintained a constant gap of 2 Å between the molecules and particles.

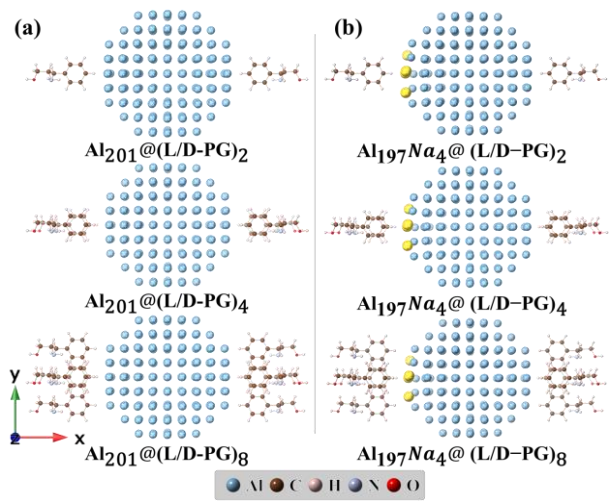

Figure 5. The geometric structures of metal cluster complexes multi-molecules. (a) the complex system for $Al_{201}@(L/D\text{-}PG)_N$ and (b) is for $Al_{197}Na_4@(L/D\text{-}PG)_N$, where $N$ stands for 2, 4, and 8.

Within the framework of LCAO-RT-TDDFT method, we obtained the photoabsorption spectra for all the systems through kick-x excitation, as shown in Figures 6a and c. We observed that as the number of molecules increases, the photoabsorption intensity of the LP significantly enhances, while the intensity of the UP diminishes. The intensity of the LP is closely related to the number of molecules; the LP intensity with 2, 4, and 8 chiral molecules adsorbed is approximately twice that of the previous, respectively. This indicates that strong coupling acts on all molecules, and their cumulative contribution leads to a significant increase in the intensity of the LP, which is closely related to more transition electrons transferring from the cluster to the molecules. More electron transitions result in a reduction of photoabsorption. Previous literature reports have indicated that molecular adsorbates can affect the LSPR of NPs, and with an increase in the number of adsorbed molecules, this effect becomes more pronounced, potentially resulting in a diminishment of the plasmonic properties[15]. Similar to the coupling of single molecule mentioned earlier, the photoabsorption spectra of multiple L-PG molecules coupled to the cluster largely coincide with those of a system where multiple D-PG molecules are coupled to the cluster. However, in terms of spectral intensity, both polaritons of $Al_{197}Na_4@(L/D\text{-}PG)_N$ are lower than those of $Al_{201}@(L/D\text{-}PG)_N$. This phenomenon could be related to the asymmetric dipole mode induced by atomic doping in the chiral cluster. This asymmetric mode hinders the formation of LSPR, resulting in a relatively weaker localized plasmonic field on the chiral cluster. Figures 6b and d list the complete spectra of all multi-molecule coupled systems, where the changes in pure molecular absorption due to the number of molecules can be observed in detail. As the number of molecules increases, the broadening of the molecular absorption becomes evident, and at the energy levels corresponding to the UP branch, the negative effect of absorption also gradually increases. This indicates a positive contribution of the increasing number of molecules to strong coupling.

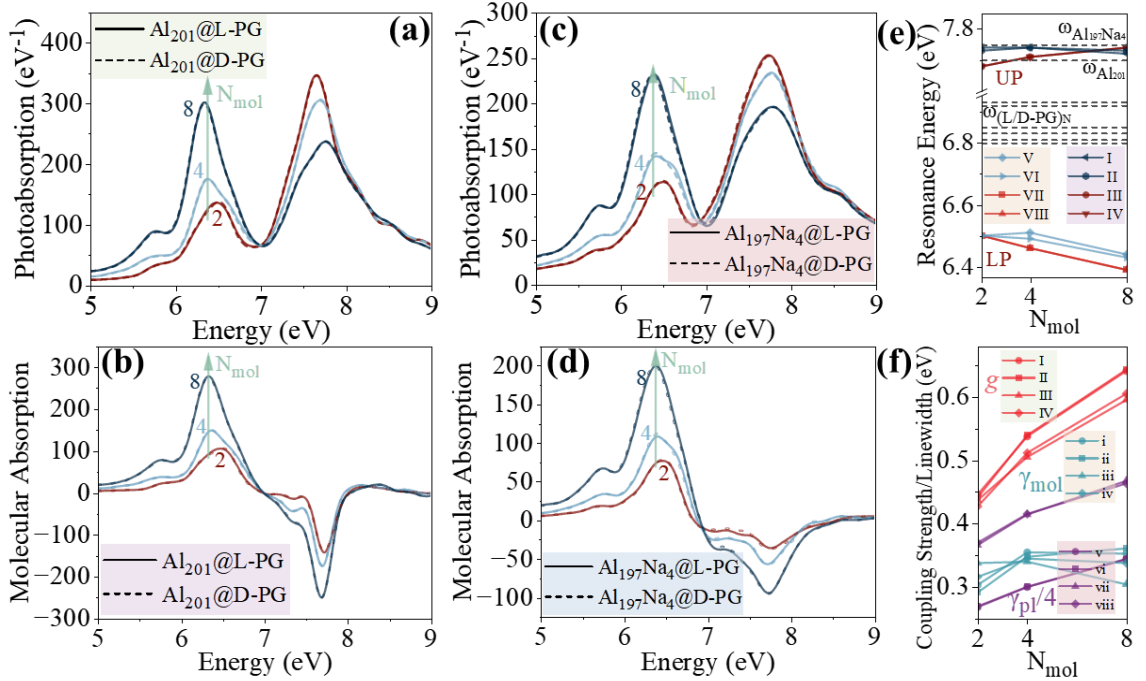

Figure 6. Photoabsorption spectrum and coupling strength for cluster complex with multi-molecular. (a) the photoabsorption for $Al_{201}@(L/D-PG)_N$, where $N$ represents the number of molecules, which are 2, 4, and 8, respectively. (b) the molecular absorption for $Al_{201}@(L/D-PG)_N$ (the total absorption spectra minus the spectrum of single cluster). (c) the photoabsorption for $Al_{197}Na_4@(L/D-PG)_N$. (d) the molecular absorption for $Al_{197}Na_4@(L/D-PG)_N$. (e) the resonance energy of $Al_{201}@(L/D-PG)_N$ and $Al_{201}@(L/D-PG)_N$, where I-IV represents the UP of $Al_{197}Na_4@(L-PG)_N$, $Al_{197}Na_4@(D-PG)_N$, $Al_{201}@(L-PG)_N$, and $Al_{201}@(D-PG)_N$, respectively, V-VIII represents the LP of $Al_{197}Na_4@(L-PG)_N$, $Al_{197}Na_4@(D-PG)_N$, $Al_{201}@(L-PG)_N$, and $Al_{201}@(D-PG)_N$, respectively. (f) the coupling strength and linewidth of $Al_{201}@(L/D-PG)_N$ and $Al_{197}Na_4@(L/D-PG)_N$ with different gap, where I-IV represents the $g$, i-iv represents the linewidth of LP, and v-viii represents the linewidth of UP for $Al_{201}@(L-PG)_N$, $Al_{201}@(D-PG)_N$, $Al_{197}Na_4@(L-PG)_N$ and $Al_{197}Na_4@(D-PG)_N$, respectively.

Plasmonic strong coupling varies with the number of coupled molecules. Clusters resonantly coupled with multi-molecules simultaneously form two stronger polaritonic states. The LP and UP are the symmetric and antisymmetric mixtures of plasmonic and molecular states, respectively. As shown in Figure 7, the two polaritonic states differ in the sign of the molecular contribution. For the LP, the plasmonic and molecular transitions are in phase, whereas for the UP, they are out of phase. These are the combinations of symmetric and antisymmetric states, which are also clearly visible in the induced densities (see Figure 7 2M, 4M, and 8M). At the LP, the dipoles of the cluster and the molecules are parallel, while at the UP, the dipoles of the cluster and the molecules are antiparallel. For $Al_{201}@(L-PG)_{2/4/8}$, as the number of coupled molecules increases, the contribution of the molecular light absorption transition significantly enhances. When the number of molecules increases to eight, the molecular contribution to photoabsorption becomes a series of strong and concentrated transitions, significantly more than those with 2 or 4 molecules. Similar to single molecule coupling systems, the LP in multi-molecule coupling systems is mainly contributed by molecular transitions, while the UP is primarily contributed by the cluster. Compared to coupling with L-PG molecules, coupling with D-PG reveals the same light absorption transition contributions and induced charge densities. For the systems with non-chiral cluster, we note that the system with two coupled molecules exhibits two transition contributions with different energies instead of one, indicating that each molecule provides a distinct transition contribution. This can also be seen from the transition intensity (the non-chiral coupled system is four times the intensity of the coupled system). Since the molecules are located on opposite sides of the cluster, and one side of the cluster is doped with *Na* atoms while the other side is not, the transition of charge to the two faces of the cluster naturally differs after coupling the molecules. This is why the TCM plot shows two distinct photoabsorption contributions with significantly different energies. Nevertheless, with an ascending number of coupled molecules, the photoabsorption contribution arising from molecular transitions progressively converges towards that observed in the non-chiral cluster system [e.g., $Al_{201}@(L-PG)_{4/8}$]. Where the doping of *Na* atoms results in an asymmetric charge distribution around the cluster. This asymmetry leads to differences in the molecular orbitals, thereby affecting the energy level splitting of the molecules. With the addition of more molecules, the subsequent molecules are located at a greater distance from the *Na* atoms and are consequently less affected by their presence. The newly incorporated molecules begin to exert a dominant influence, resulting in the TCMs of the multi-molecule systems exhibiting characteristics akin to the coupling outcomes observed with non-chiral cluster. The photoabsorption contributions of non-chiral cluster coupled with L-PG and D-PG molecules are also the same. This indicates that the chirality of the cluster significantly affects strong coupling, while the chirality isomerism of the molecules is not the main factor influencing plasmonic strong coupling. However, the chirality characteristics of chiral molecules significantly affect the system's rotary strength, which is the reason for the substantial differences in chirality across different systems. By changing the chirality of the molecules and the number of chiral molecules, we can efficiently control plasmonic strong coupling.

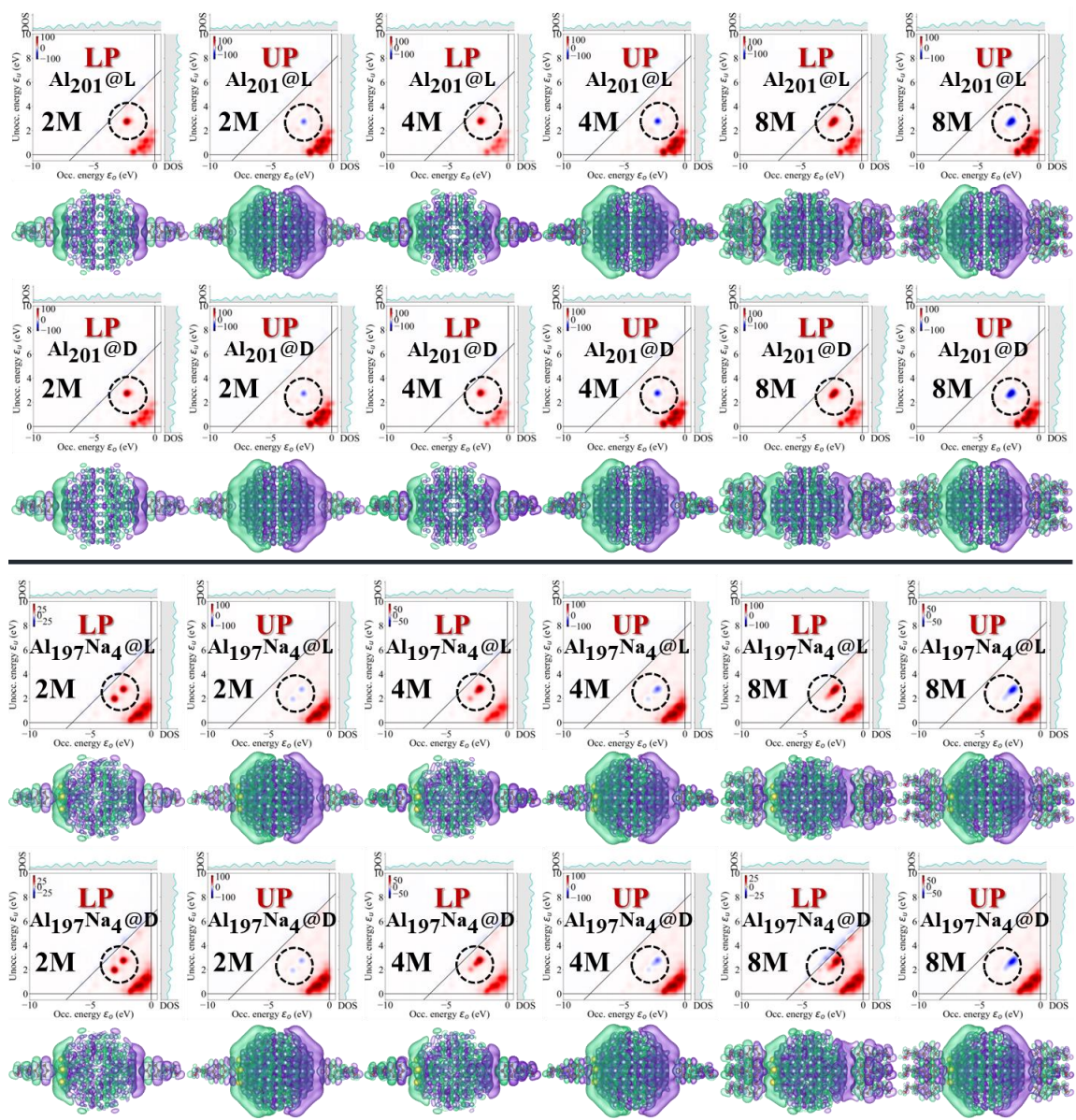

Figure 7. The TCM and induced density for $Al_{201}@(L/D-PG)_N$ and $Al_{197}Na_4@(L/D-PG)_N$, respectively, where $N$ represent 2, 4, and 8 chiral molecular. In addition, L and D represent L-PG and D-PG, respectively.

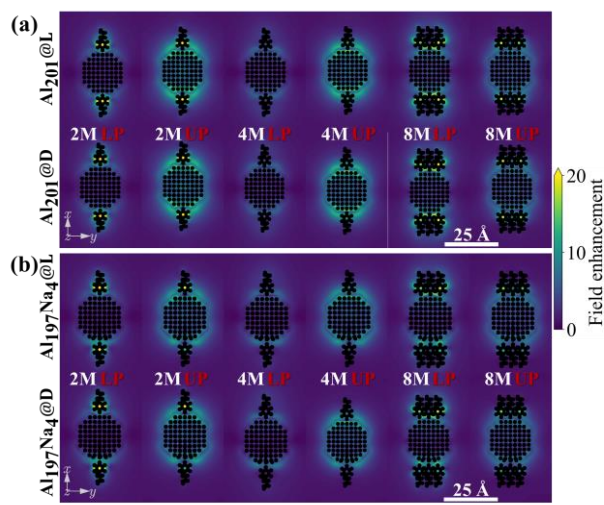

Figure 8. The electric-field enhancement of single cluster complex 2, 4, and 8 molecules, respectively. (a) the electric-field enhancement for $Al_{201}@(L-PG)_N$ and $Al_{201}@(D-PG)_N$ system, where N stands for 2, 4, and 8. (b) the electric-field enhancement for $Al_{197}Na_4@(L-PG)_N$ and $Al_{197}Na_4@(D-PG)_N$ system, where N stands for 2, 4, and 8.

Furthermore, we investigated enhancement of the electric field resulting from the simultaneous coupling of multiple chiral molecules with clusters, as shown in Figure 8. In Figure 8b, for the case of $Al_{201}@(L-PG)_2$, we observed that the enhanced LP field is primarily concentrated in the benzene ring area of the chiral molecules, which is an intrinsic field enhancement attribute of the molecule itself. At this point, the LSPR of the cluster remains unstimulated. However, for UP, we can clearly see a significant LSPR in the cluster region, with a highly enhanced field area, or "hotspot", appearing in the gap between the cluster and the molecules. This "hotspot" is the catalyst for the marked amplification of the chiral signal observed within this structure, as illustrated in Figure 9a. This indicates that the LSPR of the $Al_{201}$ cluster can effectively enhance the chirality of the L-PG molecule. As the number of adsorbed molecules increases, the chiral signal is progressively enhanced. But for $Al_{201}@(L-PG)_8$, we observed a peculiar phenomenon. For LP, the presence of numerous molecules on one side of the cluster leads to interactions between the molecules. Although the cluster's LSPR is not produced at this time, the interactions between the benzene rings of the chiral molecules result in plasmonic excitations on the molecules. This is why the field enhancement in the molecular region is so pronounced. Under such influences, the chiral signal shows a significant difference, and a substantial enhancement compared to the coupling of 2 and 4 molecules. For UP, the interaction between the cluster's LSPR and the molecular plasmons, under the combined action of LP and UP, leads to strong plasmonic coupling that greatly enhances the chirality of the molecules. Moreover, in the plasmonic enhanced chirality, we found that the molecular plasmons can also contribute sufficiently, rather than just the cluster's

LSPR participating in the excitation. For $Al_{201}@(D-PG)_N$, the field enhancement properties exhibit the same phenomena as for $(L-PG)_N$ coupling, indicating that the left and right hands of chiral molecules do not affect the mechanism of plasmonic enhanced chiral signals. The field enhancement properties of $Al_{197}Na_4@(L/D-PG)_N$ are similar, so not elaborate on this point.

The coupling strength was also quantified using the velocity-coupled harmonic oscillator model mentioned earlier, and the fitted spectra closely matched the TDDFT spectra (see Figure S7 in the *Supporting Information*). All fitting parameters are provided in Tables S5-6. Figure 6e plots the resonance energy as a function of the number of molecules. In $Al_{201}@(L/D-PG)_N$ systems, the resonance energy of the UP mode shows a blue-shift with the increasing number of molecules, and when coupled with 8 molecules, the resonance energy is already greater than the plasmon resonance energy of the isolated $Al_{201}$ cluster. In contrast, for $Al_{197}Na_4@(L/D-PG)_N$ systems, the resonance energy of the UP mode red-shifts with the increasing number of molecules, and when coupled with 8 molecules, the resonance energy is already less than the resonance energy of the isolated $Al_{197}Na_4$ cluster. However, for the LP mode of the coupled systems, a pronounced red-shift in resonance energy is observed with the increasing number of molecules, with $Al_{201}@(L/D-PG)_N$ systems exhibiting a stronger red-shift. Figure 6f summarizes the resonance peak broadening and coupling strength for both systems. It is evident that the coupling strength increases with the number of molecules coupled to the cluster, with the $Al_{201}@(L/D-PG)_N$ systems being stronger than the $Al_{197}Na_4@(L/D-PG)_N$ systems. This is consistent with the stronger negative absorption effect observed in Figure 6b. Additionally, the fitting width indicates a significant increase in the oscillator width, with the multi-molecule coupled systems having a pronounced effect on the UP absorption.

**Enhanced Chirality by Molecular Collective Resonance.** Driven by the unexpected strong coupling and pronounced Rabi splitting observed in the multi-molecule systems, we have developed a further interest in the mechanism behind the enhanced chirality. To facilitate the calculation of the chirality enhancement factor, we additionally computed the absorption spectra of the pure molecular systems after the removal of the clusters. As seen in Figure S8 of the *Supporting Information*, the absorption spectra exhibit a gradual blue-shift with an increase in the number of molecules, along with an enhancement in absorption intensity. Based on this, we calculated the rotary strengths for both the cluster-coupled multi-molecular systems and the pure molecular systems, as depicted in Figure 9a-c and Figure S9.

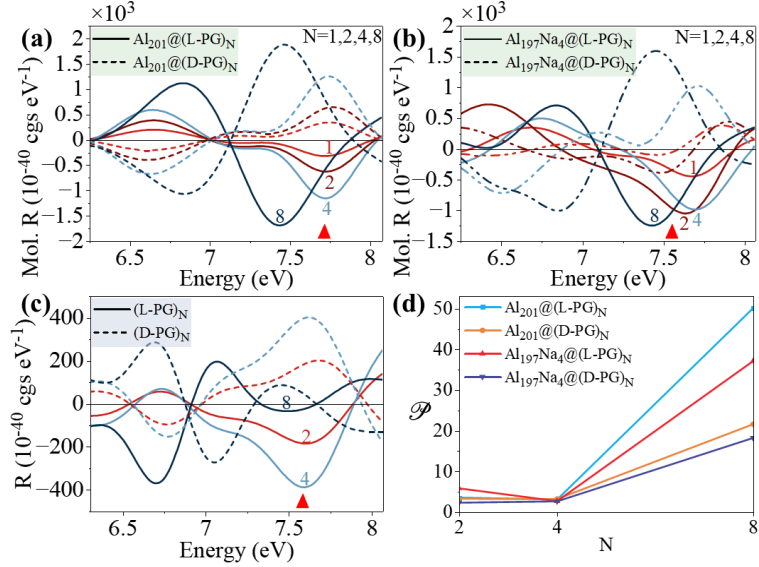

Figure 9. The molecular rotor strength. The CD spectra of (a) $Al_{201}$ and (b) $Al_{197}Na_4$ complexes 2, 4, and 8 L/D-PG chiral molecular, respectively. (c) is the CD spectrum of all pure molecular system. The red triangle in the figure marks the position of the peak of interest. (d) is the enhanced chirality factor.

For systems involving the coupling of multiple molecules to non-chiral cluster, we observe more regular CD spectral lines, and the CD intensity increases with the number of coupled molecules (Figure 9a). In the case of $Al_{201}$@(L/D-PG)$_{2/4}$ systems, the CD peak positions of interest remain consistent, but the CD intensity for N=4 is approximately twice that of N=2, which is also roughly twice the intensity of N=1, providing a clear comparison. Hence, when N≤4, the enhancement of chirality is directly proportional to the number of molecules added, leading to a nearly constant enhancement factor (Figure 9d), as the CD intensity of the pure molecules also exhibits an approximately twofold increase (Figure 9c). However, at N=8, we observe a significant deviation in CD properties compared to N≤4. Notably, the CD peak positions of interest experience a marked red-shift, and the intensity is significantly enhanced. In contrast, the CD intensity of the pure molecules is even lower than that for N≤4, which accounts for the larger value of the chirality enhancement factor observed (Figure 9d). In summary, when the count of molecules is N≤4, the gap between the molecules across both sides of the cluster is considerable, leading to negligible intermolecular interactions, even in the presence of two molecules on a single side. Consequently, as the number of coupled molecules increases, the coupling strength exhibits a simple linear superposition, which does not significantly contribute to the chirality enhancement factor. In this regime, the chirality enhancement is primarily driven by the cluster plasmon-induced enhancement of molecular chirality (resonant interactions). However, when the number of molecules increases to N=8, with four molecules on one side of the cluster, significant intermolecular interactions become evident. This results in a profound differentiation in the chiral signal of the pure molecular system, ultimately leading to a dramatic increase in the chirality enhancement factor. At this stage, there appears to be an additional mechanism for chirality enhancement, namely the enhancement of chirality due to molecule-molecule interactions (non-resonant interactions).

For systems with chiral clusters, the CD spectra exhibit pronounced discordance (Figure 9b). For N≤4, although the spectra are relatively disordered, the chirality enhancement factor calculated from the peaks of interest reveals that, within the margin of computational error, it maintains the same features as the non-chiral cluster systems remaining almost constant. When N=8, the chirality enhancement factor similarly experiences a sharp increase, further substantiating the significant impact of intermolecular interactions on the chiral properties. This is corroborated by the induced charge densities (Figure 7). Additionally, we note that the chirality enhancement factor in systems with chiral cluster is weaker compared to those with non-chiral cluster. This indicates that the inherent chirality of the cluster itself also influences the overall enhancement, suggesting a third chiral mechanism: modal crosstalk or resonance between NP chirality and molecular chirality.

**Conclusions**

This work provides a comprehensive investigation into molecular chirality enhancement and coupling dichroism in strongly coupled chiral systems at the atomic level. Utilizing RT-TDDFT simulations, we have explored the influence of key factors such as the cluster-molecule gap and the number of coupled molecules on the enhancement of CD signals. We have demonstrated that strong plasmon-molecule coupling leads to the formation of hybridized polarization modes, whose behavior is dictated by the geometric arrangement and chirality of both clusters and molecules. Na-doped Al cluster introduces intrinsic chiral characteristics, further enhancing the molecular CD response, particularly at small gaps. This study highlights the critical role of gap size in controlling the coupling strength and CD signal enhancement and reveals both of the coupling factor and decay rate of the coupled system will be modulated by the chirality of the molecules and the cluster. Moreover, the observed synergistic enhancement effects in multi-molecule systems indicate that the enhanced chirality is linearly influenced by the number of molecules when the number of coupled molecules is N≤4, with their enhancement factors consistent with the single-molecule case. Therefore, high-efficiency NPs that facilitate molecular adsorption should be selected in the construction of molecular chirality sensors. When the number of coupled molecules reaches 8, a significant deviation is observed, with a red shift in CD peak positions and a substantial increase in intensity, suggesting the presence of stronger intermolecular interactions and a new mechanism for chirality enhancement beyond the cluster plasmonic induced effects, the coupled molecules leads to a substantial increase in the intensity of lower polaritonic modes, highlighting the collective behavior in multi-molecule systems due to the modal crosstalk or resonance between cluster chirality and molecular chirality.Our work systematically investigates the impact of plasmonic enhancement on the circular dichroism of chiral molecules, uncovering the microscopic mechanisms of strongly coupled chiral molecule-plasmonic NP systems, and provides important theoretical foundations for research and applications in the field of chiral optics.

**Authors contributions**
Y.F. conceived the idea and directed the project. N.G. and H.L. performed the RT-TDDFT calculations and data analysis. N.G., H.L. and Y.F. planned the research and wrote the paper. All of the authors revised the manuscript.

**Funding**
This research was supported by the National Natural Science Foundation of China (Grant No. 12274054).

**Conflicts of interest**
The authors declare no competing financial interest.

**Availability of data and material**
The data and material that support the findings of this study are available from the corresponding author upon reasonable request.

Supporting Information for
**Coupling Dichroism in Strong-Coupled Chiral Molecule-Plasmon Nanoparticle System**

Nan Gao[1, #], Haoran Liu[1, #], Yurui Fang[1, *]

[1.] School of Physics, Dalian University of Technology, Dalian 116024, P.R. China.

* Corresponding authors: <u>yrfang@dlut.edu.cn</u> (Y.F.)

# These authors contributed equally.

**Figures.**

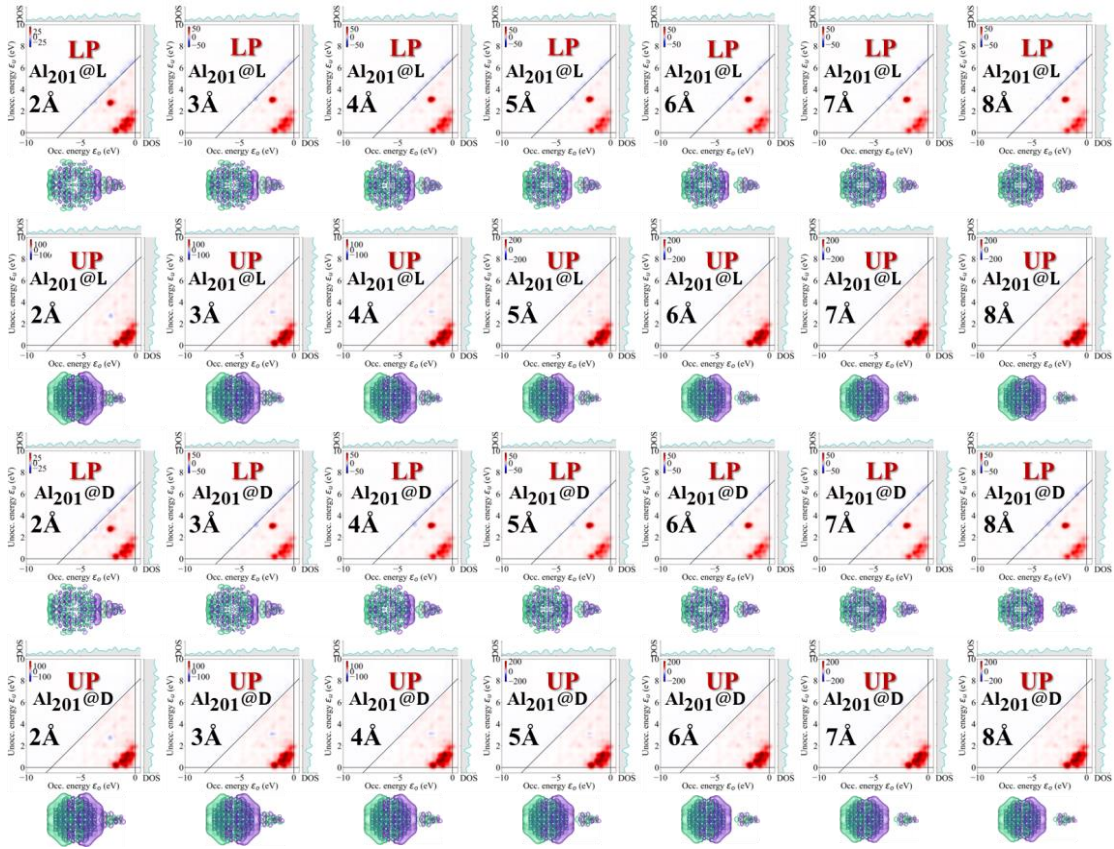

Figure S1. The TCM and induced density for $Al_{201}$@L-PG and $Al_{201}$@D-PG with different distance, the distance is 2-8 Å.

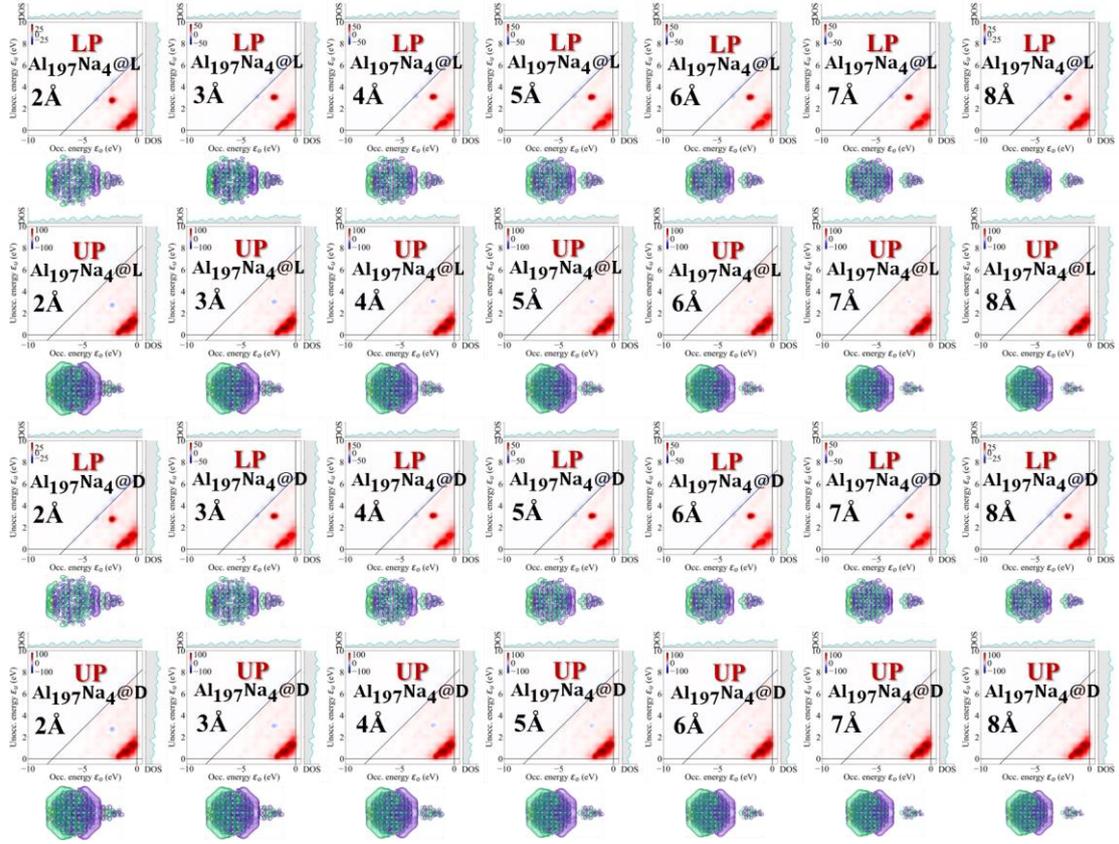

Figure S2. The TCM and induced density for $Al_{197}Na_4$@L-PG and $Al_{197}Na_4$@D-PG with different distance, the distance is 2-8 Å.

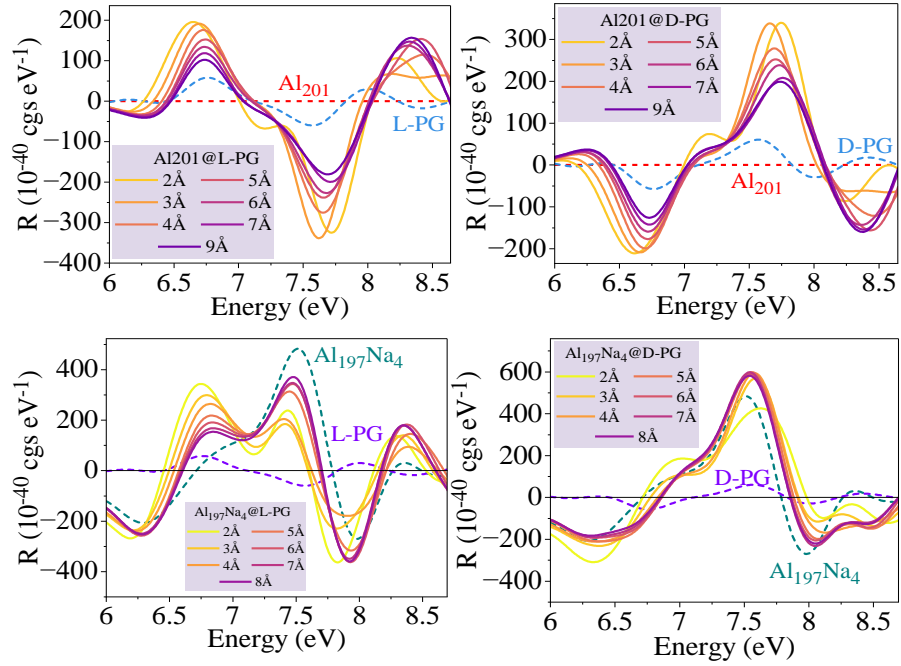

Figure S3. The plasmon enhanced CD spectra for total system with different gap sizes.

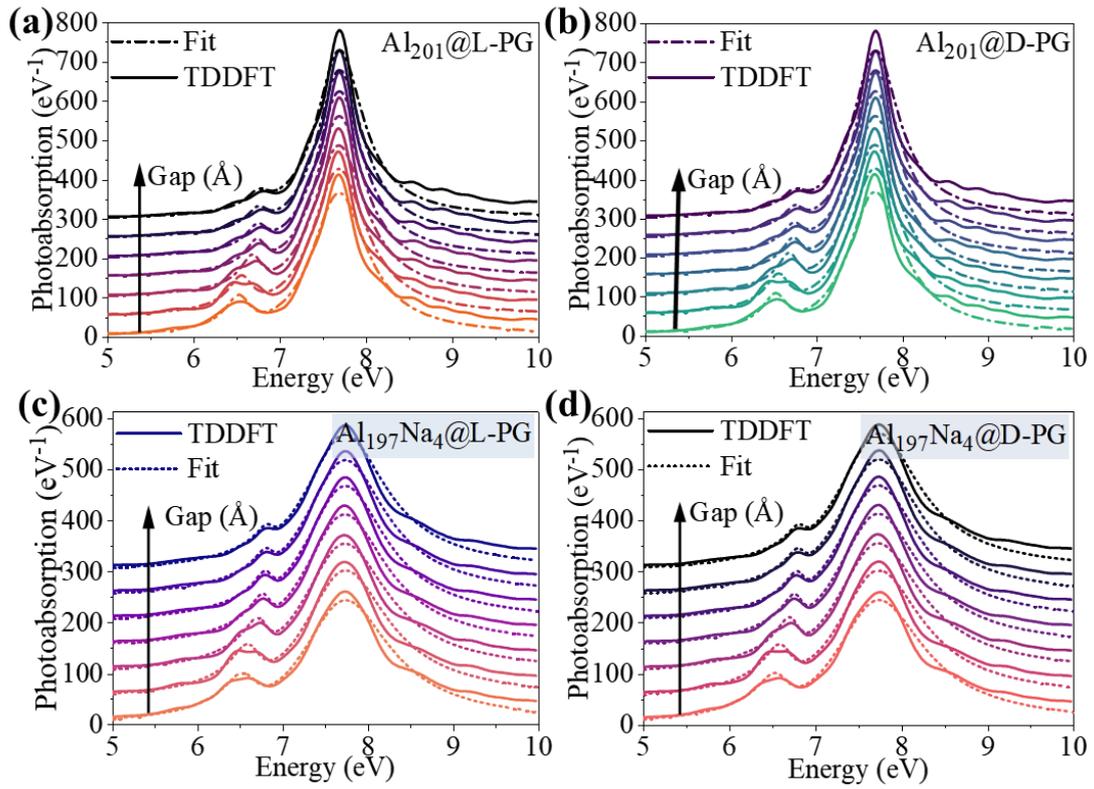

Figure S4. Fitted spectral by Coupled Oscillator Model (COM). (a) the LCAO-RT-TDDFT spectrum vs fitting spectrum for $Al_{201}$@L-PG with 2-8 Å. (b) the LCAO-RT-TDDFT spectrum vs fitting spectrum for $Al_{201}$@D-PG with 2-8 Å. (c) the LCAO-RT-TDDFT spectrum vs fitting spectrum for $Al_{197}Na_4$@L-P with 2-8 Å. (d) the LCAO-RT-TDDFT spectrum vs fitting spectrum for $Al_{197}Na_4$@D-PG with 2-8 Å.

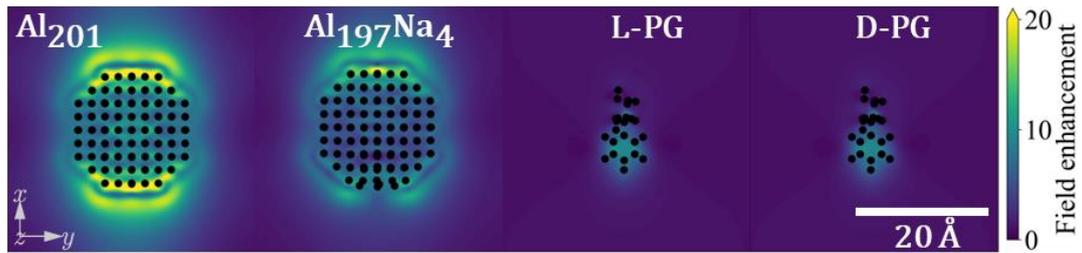

Figure S5. The electric-field enhancement for single system ($Al_{201}$, $Al_{197}Na_4$, L-PG, and D-PG).

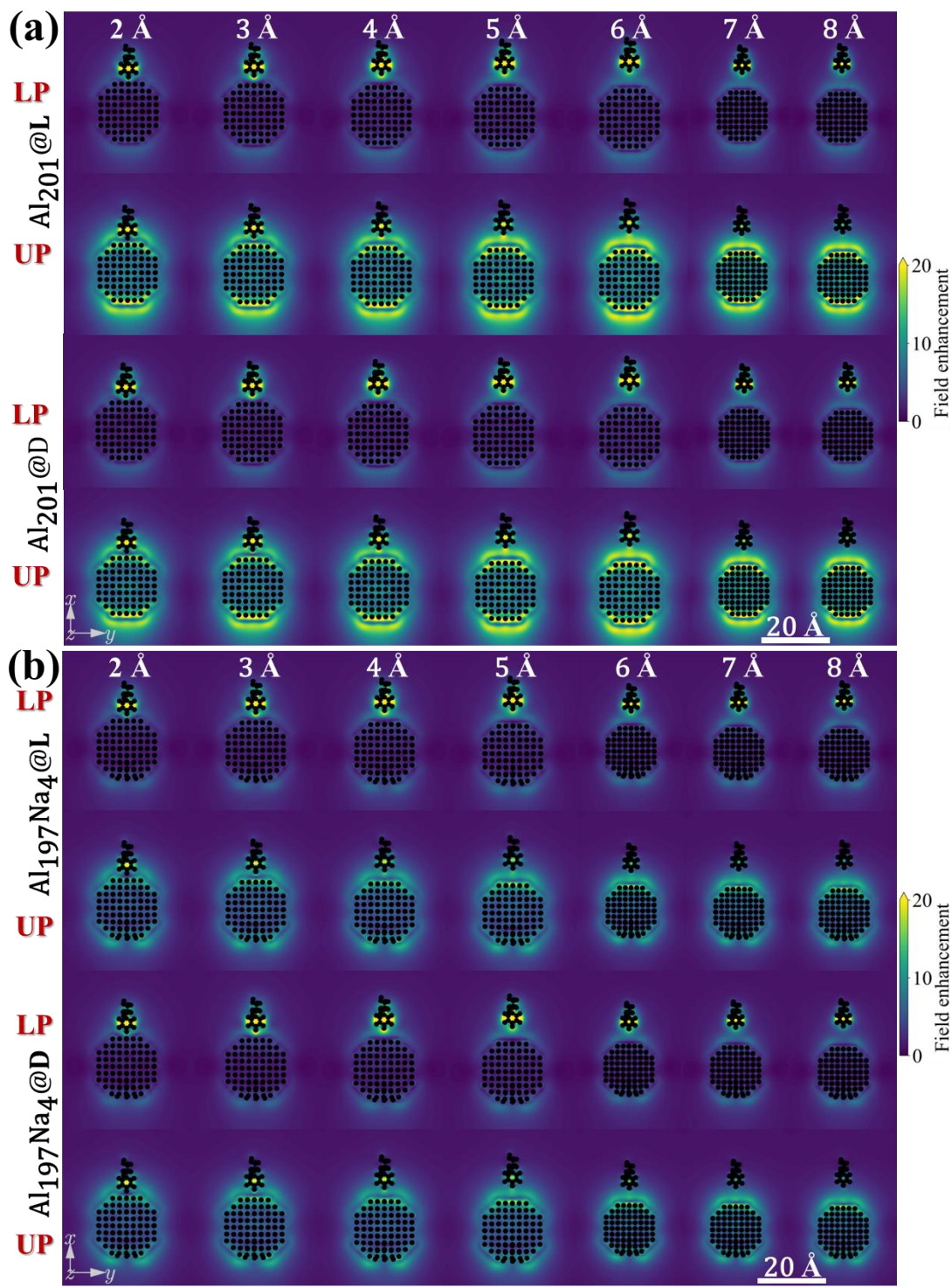

Figure S6. The electrical field enhancement at different gap between molecules and clusters. (a-b) the electrical field for $Al_{201}$@L/D-PG and $Al_{197}Na_4$@L/D-PG with different gap, respectively.

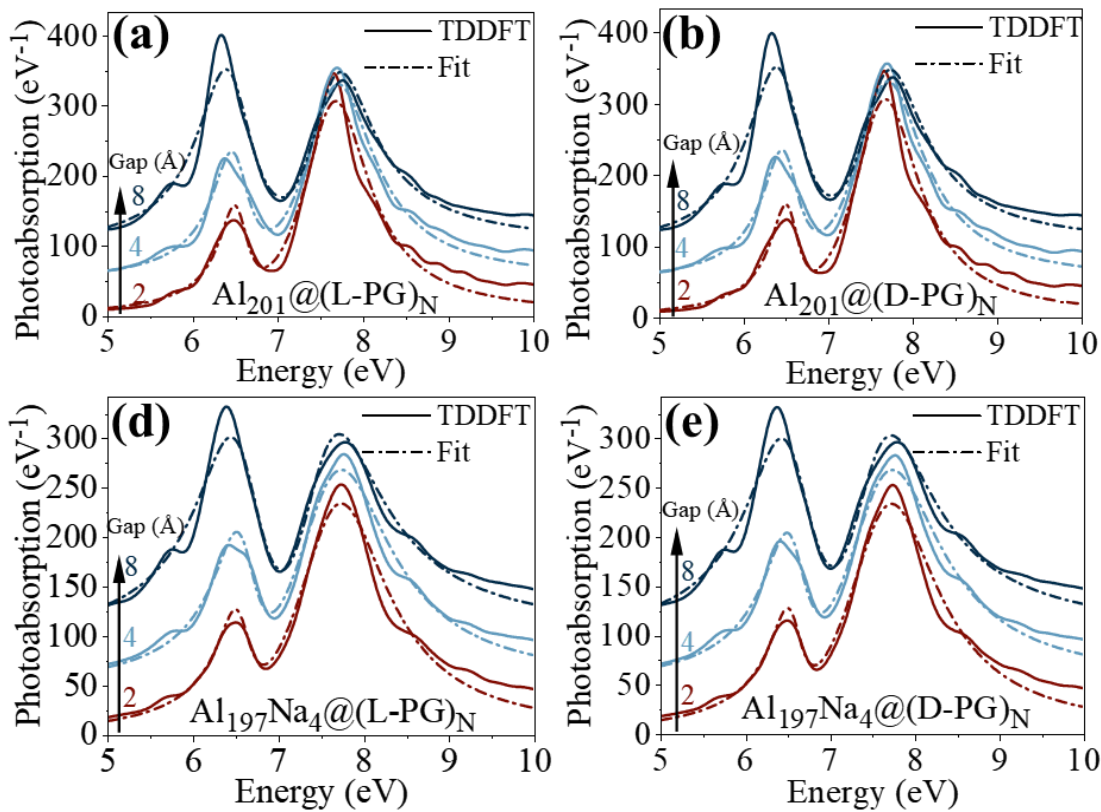

Figure S7. Fitted spectral by COM. (a) the LCAO-RT-TDDFT spectrum vs fitting spectrum for Al$_{201}$@(L-PG)$_N$ (N represent 2, 4, and 8). (b) the LCAO-RT-TDDFT spectrum vs fitting spectrum for Al$_{201}$@(D-PG)$_N$ (N represent 2, 4, and 8). (c) the LCAO-RT-TDDFT spectrum vs fitting spectrum for Al$_{197}$Na$_4$@(L-PG)$_N$ (N represent 2, 4, and 8). (d) the LCAO-RT-TDDFT spectrum vs fitting spectrum for Al$_{197}$Na$_4$@(D-PG)$_N$ (N represent 2, 4, and 8).

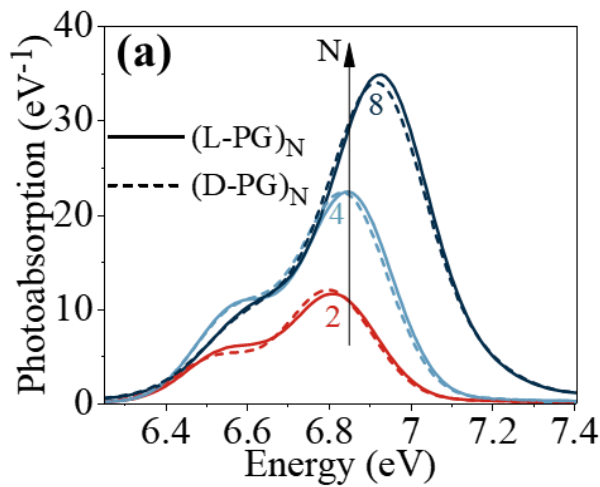

Figure S8. The photoabsorption spectrum of (L/D-PG)$_N$, where N is 2, 4, and 8, respectively.

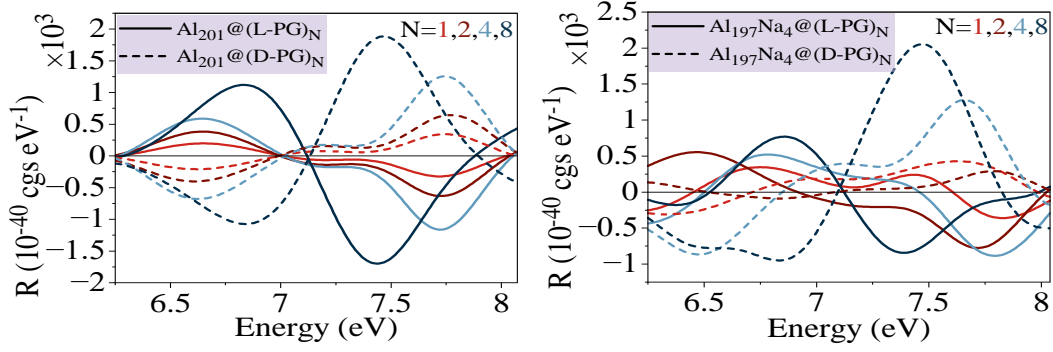

Figure S9. The total system rotor strength for coupled cluster and different number of molecules.

**Tables.**

Table S1. The fitting parameters of velocity oscillator model for fitting for $Al_{201}$@L-PG with different.

| $d$ (Å) | $g$ (eV) | $a$ (1/eV) | $\omega_{ex}$ (eV) | $\gamma_{ex}$ (eV) | $\omega_{pl}$ (eV) | $\gamma_{pl}$ (eV) |
|---|---|---|---|---|---|---|
| 2.0 | 0.352 | 315.897 | 6.53 | 0.320 | 7.69 | 0.822 |
| 3.0 | 0.335 | 312.744 | 6.56 | 0.309 | 7.69 | 0.792 |
| 4.0 | 0.272 | 309.572 | 6.69 | 0.262 | 7.69 | 0.775 |
| 5.0 | 0.236 | 302.555 | 6.74 | 0.261 | 7.69 | 0.715 |
| 6.0 | 0.218 | 299.182 | 6.75 | 0.275 | 7.70 | 0.686 |
| 7.0 | 0.204 | 297.690 | 6.77 | 0.283 | 7.70 | 0.678 |
| 8.0 | 0.191 | 296.843 | 6.78 | 0.289 | 7.70 | 0.676 |

Table S2. The fitting parameters of velocity oscillator model for fitting for $Al_{201}$@D-PG with different.

| $d$ (Å) | $g$ (eV) | $a$ (1/eV) | $\omega_{ex}$ (eV) | $\gamma_{ex}$ (eV) | $\omega_{pl}$ (eV) | $\gamma_{pl}$ (eV) |
|---|---|---|---|---|---|---|
| 2.0 | 0.348 | 315.889 | 6.53 | 0.310 | 7.69 | 0.823 |
| 3.0 | 0.334 | 312.763 | 6.56 | 0.304 | 7.69 | 0.793 |
| 4.0 | 0.270 | 309.506 | 6.69 | 0.251 | 7.69 | 0.775 |
| 5.0 | 0.236 | 302.698 | 6.74 | 0.254 | 7.69 | 0.716 |
| 6.0 | 0.218 | 299.184 | 6.75 | 0.269 | 7.70 | 0.686 |
| 7.0 | 0.203 | 297.577 | 6.77 | 0.274 | 7.70 | 0.678 |
| 8.0 | 0.191 | 296.703 | 6.78 | 0.285 | 7.70 | 0.676 |

Table S3. The fitting parameters of velocity oscillator model for fitting for $Al_{197}Na_4$@L-PG with different.

| $d$ (Å) | $g$ (eV) | $a$ (1/eV) | $\omega_{ex}$ (eV) | $\gamma_{ex}$ (eV) | $\omega_{pl}$ (eV) | $\gamma_{pl}$ (eV) |
|---|---|---|---|---|---|---|
| 2.0 | 0.367 | 342.865 | 6.54 | 0.393 | 7.75 | 1.356 |
| 3.0 | 0.346 | 336.681 | 6.59 | 0.339 | 7.75 | 1.292 |
| 4.0 | 0.269 | 334.794 | 6.71 | 0.266 | 7.74 | 1.288 |
| 5.0 | 0.218 | 330.785 | 6.77 | 0.230 | 7.74 | 1.245 |
| 6.0 | 0.190 | 327.324 | 6.80 | 0.219 | 7.74 | 1.209 |
| 7.0 | 0.170 | 325.915 | 6.82 | 0.222 | 7.74 | 1.200 |
| 8.0 | 0.154 | 324.964 | 6.83 | 0.221 | 7.74 | 1.196 |

Table S4. The fitting parameters of velocity oscillator model for fitting for $Al_{197}Na_4$@D-PG with different.

| $d$ (Å) | $g$ (eV) | $a$ (1/eV) | $\omega_{ex}$ (eV) | $\gamma_{ex}$ (eV) | $\omega_{pl}$ (eV) | $\gamma_{pl}$ (eV) |
|---|---|---|---|---|---|---|
| 2.0 | 0.365 | 342.975 | 6.54 | 0.383 | 7.75 | 1.360 |
| 3.0 | 0.344 | 336.856 | 6.59 | 0.329 | 7.75 | 1.294 |
| 4.0 | 0.271 | 334.777 | 6.71 | 0.260 | 7.74 | 1.287 |
| 5.0 | 0.220 | 330.705 | 6.77 | 0.225 | 7.74 | 1.244 |
| 6.0 | 0.192 | 327.367 | 6.80 | 0.218 | 7.74 | 1.208 |
| 7.0 | 0.172 | 325.961 | 6.82 | 0.222 | 7.74 | 1.200 |
| 8.0 | 0.157 | 325.063 | 6.83 | 0.222 | 7.74 | 1.196 |

Table S5. The fitting parameters of velocity oscillator model for fitting for $Al_{201}$@(L/D-PG)$_N$.

| Num. | $g$ (eV) | $a$ (1/eV) | $\omega_{ex}$ (eV) | $\gamma_{ex}$ (eV) | $\omega_{pl}$ (eV) | $\gamma_{pl}$ (eV) |
|---|---|---|---|---|---|---|
| (L-PG)$_2$ | 0.446 | 350.234 | 6.50 | 0.302 | 7.68 | 1.068 |
| (L-PG)$_4$ | 0.538 | 375.002 | 6.46 | 0.354 | 7.71 | 1.196 |
| (L-PG)$_8$ | 0.642 | 425.232 | 6.39 | 0.352 | 7.74 | 1.375 |
| (D-PG)$_2$ | 0.441 | 350.345 | 6.50 | 0.290 | 7.68 | 1.073 |
| (D-PG)$_4$ | 0.540 | 375.316 | 6.46 | 0.347 | 7.71 | 1.198 |
| (D-PG)$_8$ | 0.644 | 424.993 | 6.39 | 0.360 | 7.74 | 1.372 |

Table S6. The fitting parameters of velocity oscillator model for fitting for $Al_{197}Na_4$@(L/D-PG)$_N$.

| Num. | $g$ (eV) | $a$ (1/eV) | $\omega_{ex}$ (eV) | $\gamma_{ex}$ (eV) | $\omega_{pl}$ (eV) | $\gamma_{pl}$ (eV) |
|---|---|---|---|---|---|---|
| (L-PG)$_2$ | 0.437 | 358.380 | 6.50 | 0.337 | 7.74 | 1.462 |
| (L-PG)$_4$ | 0.505 | 386.713 | 6.51 | 0.339 | 7.74 | 1.656 |
| (L-PG)$_8$ | 0.596 | 434.939 | 6.44 | 0.303 | 7.72 | 1.870 |
| (D-PG)$_2$ | 0.427 | 359.205 | 6.50 | 0.315 | 7.73 | 1.478 |
| (D-PG)$_4$ | 0.512 | 387.225 | 6.49 | 0.344 | 7.74 | 1.660 |
| (D-PG)$_8$ | 0.606 | 434.050 | 6.43 | 0.337 | 7.73 | 1.853 |

Table S7. The enhancement chirality factor for $Al_{201}$@L-PG with different gap. Where I and II represent the first positive (negative) peak or the second negative (positive) peak, respectively.

| Gap | I | II | $\mathscr{P}^{\text{I}}$ | $\mathscr{P}^{\text{II}}$ |
|---|---|---|---|---|
| 2 Å | 195.46 | -325.65 | 3.38 | 5.41 |
| 3 Å | 191.25 | -338.96 | 3.31 | 5.63 |
| 4 Å | 175.76 | -275.83 | 3.04 | 4.58 |
| 5 Å | 152.23 | -238.79 | 2.63 | 3.97 |
| 6 Å | 133.93 | -227.44 | 2.32 | 3.78 |
| 7 Å | 118.40 | -199.74 | 2.05 | 3.32 |
| 8 Å | 102.03 | -181.25 | 1.77 | 3.01 |

Table S8. The enhancement chirality factor for $Al_{201}$@D-PG with different gap. Where I and II represent the first positive (negative) peak or the second negative (positive) peak, respectively.

| Gap | I | II | $\mathscr{P}^{I}$ | $\mathscr{P}^{II}$ |
|---|---|---|---|---|
| 2 Å | -211.46 | 339.85 | 3.66 | 5.64 |
| 3 Å | -209.09 | 338.44 | 3.62 | 5.62 |
| 4 Å | -199.53 | 278.33 | 3.45 | 4.62 |
| 5 Å | -177.42 | 252.18 | 3.07 | 4.19 |
| 6 Å | -159.51 | 238.00 | 2.76 | 3.95 |
| 7 Å | -142.75 | 207.98 | 2.47 | 3.45 |
| 8 Å | -126.37 | 198.98 | 2.19 | 3.30 |

Table S9. The enhancement chirality factor for $Al_{197}Na_4$@L-PG with different gap. Where I and II represent the first positive (negative) peak or the second negative (positive) peak, respectively.

| Gap | I | II | $\mathscr{P}^{I}$ | $\mathscr{P}^{II}$ |
|---|---|---|---|---|
| 2 Å | 341.15 | -455.23 | 5.90 | 7.56 |
| 3 Å | 273.77 | -438.43 | 4.74 | 7.28 |
| 4 Å | 225.23 | -391.92 | 3.90 | 6.51 |
| 5 Å | 174.45 | -269.65 | 3.02 | 4.48 |
| 6 Å | 146.01 | -227.54 | 2.53 | 3.78 |
| 7 Å | 123.27 | -229.51 | 2.13 | 3.81 |
| 8 Å | 110.04 | -213.03 | 1.90 | 3.54 |

Table S10. The enhancement chirality factor for $Al_{197}Na_4$@D-PG with different gap. Where I and II represent the first positive (negative) peak or the second negative (positive) peak, respectively.

| Gap | I | II | $\mathscr{P}^{I}$ | $\mathscr{P}^{II}$ |
|---|---|---|---|---|
| 2 Å | -114.08 | 376.66 | 1.97 | 6.25 |
| 3 Å | -85.53 | 291.00 | 1.48 | 4.83 |
| 4 Å | -99.03 | 276.27 | 1.71 | 4.59 |
| 5 Å | -128.80 | 233.10 | 2.23 | 3.87 |
| 6 Å | -122.42 | 189.84 | 2.12 | 3.15 |
| 7 Å | -110.96 | 183.33 | 1.92 | 3.04 |
| 8 Å | -101.25 | 174.25 | 1.75 | 2.89 |

Table S11. The enhancement chirality factor for $Al_{201}@(L/D\text{-}PG)_N$ and $Al_{197}Na_4@(L/D\text{-}PG)_N$.

|   | Peak |   | Peak | $\mathcal{P}$ |
|---|---|---|---|---|
| $Al_{201}@(L\text{-}PG)_2$ | -634.94 | $(L\text{-}PG)_2$ | -184.88 | 3.43 |
| $Al_{201}@(L\text{-}PG)_4$ | -1164.32 | $(L\text{-}PG)_4$ | -388.61 | 3.00 |
| $Al_{201}@(L\text{-}PG)_8$ | -1701.93 | $(L\text{-}PG)_8$ | -33.90 | 50.20 |
| $Al_{201}@(D\text{-}PG)_2$ | 644.44 | $(D\text{-}PG)_2$ | 202.48 | 3.18 |
| $Al_{201}@(D\text{-}PG)_4$ | 1253.29 | $(D\text{-}PG)_4$ | 403.55 | 3.11 |
| $Al_{201}@(D\text{-}PG)_8$ | 1884.04 | $(D\text{-}PG)_8$ | 87.41 | 21.55 |
| $Al_{197}Na_4@(L\text{-}PG)_2$ | -1060.97 | $(L\text{-}PG)_2$ | -184.88 | 5.74 |
| $Al_{197}Na_4@(L\text{-}PG)_4$ | -995.38 | $(L\text{-}PG)_4$ | -388.61 | 2.56 |
| $Al_{197}Na_4@(L\text{-}PG)_8$ | -1258.70 | $(L\text{-}PG)_8$ | -33.90 | 37.13 |
| $Al_{197}Na_4@(D\text{-}PG)_2$ | 436.52 | $(D\text{-}PG)_2$ | 202.48 | 2.16 |
| $Al_{197}Na_4@(D\text{-}PG)_4$ | 1018.87 | $(D\text{-}PG)_4$ | 403.55 | 2.52 |
| $Al_{197}Na_4@(D\text{-}PG)_8$ | 1589.90 | $(D\text{-}PG)_8$ | 87.41 | 18.19 |

**Supplementary Notes:**
**Supplementary Notes S1: Transition Contribution Maps.**
To facilitate the examination of responses according to the Kohn-Sham analysis, we utilize a transition contribution map (TCM) approach.[1,2] This method is particularly beneficial for studying plasmonic systems, where the resonances are usually composed of numerous electron-hole pair transitions. The TCM depicts the Kohn-Sham decomposition weights $w_{ia}(\omega)$ at a constant frequency ω within a bi-dimensional space, referred to as the $(\varepsilon_o, \varepsilon_u)$-plane. This plane is characterized by the energy coordinates of the filled and empty electronic states. In essence, the 2D graphical representation is constructed based on the weights of these transitions:

$$M_\omega^{\text{TCM}}(\varepsilon_o, \varepsilon_u) = \sum_{ia} w_{ia}(\omega) g_{ia}(\varepsilon_o, \varepsilon_u) \tag{1.1}$$

where $g_{ia}$ denotes a two-dimensional smoothing kernel applied to the discrete Kohn-Sham $i \rightarrow a$ transition contributions. For this purpose, we utilize a two-dimensional Gaussian distribution, denoted as:

$$g_{ia}(\varepsilon_o, \varepsilon_u) = \frac{1}{2\pi\sigma^2} \exp\left[-\frac{(\varepsilon_o - \epsilon_i)^2 + (\varepsilon_u - \epsilon_a)^2}{2\sigma^2}\right] \tag{1.2}$$

apply a Gaussian smoothing kernel with a standard deviation of σ=0.1 eV to assign a tangible extent to each $i \rightarrow a$ contribution. This identical σ value is maintained for the purpose of spectral broadening. Regarding the weight $w_{ia}(\omega)$, it is computed from the absorption decomposition, normalized relative to the overall absorption, expressed as:

$$w_{ia}(\omega) = S_{ia}^x(\omega)/S_x(\omega) \tag{1.3}$$

where KS decomposition of the absorption spectrum as

$$S_{ia}^x(\omega) = -\frac{4\omega}{\pi} \text{Im}[\mu_{ia}^{x*} \delta\rho_{ia}^x(\omega)] \tag{1.4}$$

Analyses akin to photoabsorption decomposition have been conducted within the electron-hole domain, utilizing criteria such as spatial coordinates or angular momentum. After identifying the pertinent Kohn-Sham transitions associated with a particular resonance, one can examine the real-space density changes they elicit in the resonance. The density alterations stemming from the transition $i \rightarrow a$ are quantified as:

$$\delta n_{ia}^z(\mathbf{r}, \omega) = 2\psi_i^{(0)}(\mathbf{r}) \psi_a^{(0)*}(\mathbf{r}) \delta\rho_{ia}^z(\omega) \tag{1.5}$$